\documentclass{aastex62}
\pdfoutput=1

\received{May 23, 2018}
\revised{June 13, 2018}
\accepted{June 19, 2018}

\submitjournal{ApJL}

\shorttitle{A Gap in Main Sequence} 
\shortauthors{Jao et al.}

\begin{document}

\title{A Gap in the Lower Main Sequence Revealed by
    {\it Gaia} Data Release 2}

\correspondingauthor{Wei-Chun Jao}
\email{jao@astro.gsu.edu}

\author[0000-0003-0193-2187]{Wei-Chun Jao}
\affil{Department of Physics and Astronomy \\
Georgia State University \\
Atlanta, GA 30303, USA}

\author{Todd J. Henry}
\affil{RECONS Institute\\ 
Chambersburg, PA 17201, USA}

\author{Douglas R. Gies}
\affil{Center for High Angular Resolution Astronomy and 
Department of Physics and Astronomy\\
Georgia State University\\
P.O. Box 5060, Atlanta, GA 30302-5060, USA}

\author{Nigel C. Hambly} 
\affil{Institute for Astronomy, School of Physics and Astronomy\\
University of Edinburgh, Royal Observatory\\
Blackford Hill, Edinburgh, EH9 3HJ, UK}


\begin{abstract}

  We present the discovery of a gap near $M_{G}\approx$10 in the main
  sequence on the Hertzsprung-Russell Diagram (HRD) based on
  measurements presented in {\it Gaia} Data Release 2 (DR2).  Using an
  observational form of the HRD with $M_{G}$ representing luminosity
  and $G_{BP}-G_{RP}$ representing temperature, the gap presents a
  diagonal feature that dips toward lower luminosities at redder
  colors.  The gap is seen in samples extracted from DR2 with various
  distances, and is not unique to the {\it Gaia} photometry --- it
  also appears when using near-IR photometry ($J-K_{s}$ vs
  $M_{K_{s}}$).  The gap is very narrow ($\sim$0.05 mag) and is near
  the luminosity-temperature regime where M dwarf stars transition
  from partially to fully convective, i.e., near spectral type M3.0V.
  This gap provides a new feature in the H-R Diagram that hints at an
  underlying astrophysical cause and we propose that it is linked to
  the onset of full convection in M dwarfs.

\end{abstract}



\section{Introduction} 
\label{sec:intro}

The Hertzsprung-Russell Diagram (HRD) provides the most fundamental
map in stellar astronomy, relating key information about stars of many
types.  The HRD typically relates a star's luminosity and temperature,
but we can also derive related quantities that change a star's precise
location on the HRD, including its size, metallicity, and of course,
evolutionary state.  To place a star on an HRD, astronomers need an
essential measurement of its distance, most reliably measured via a
trigonometric parallax.  The practice of measuring parallaxes started
with F.~Bessel, who determined the parallax for 61 Cygni in 1838
\citep{Bessel1838}.  Since then, astronomers have made most parallax
measurements from the ground by pointing telescopes at stars one at a
time.  Results of ground-based astrometry programs as of 1995 were
summarized in {\it The General Catalogue of Trigonometric Stellar
  Parallaxes} (also known as the ``Yale Parallax Catalog'', or YPC)
\citep{YPC}, a valuable compendium to which more recent programs have
added.

In 1989, the European Space Agency (ESA) launched the {\it Hipparcos}
mission to measure parallaxes systematically for most stars brighter
than $V\approx$9, and improved the precision on parallaxes to a few
milli-arcseconds in most cases \citep{Hipparcos}.  Although
three-dimensional space was well-mapped by {\it Hipparcos} for bright
stars, distances for relatively few of the ubiquitous red dwarfs were
measured.  Although the YPC included parallaxes for several thousand
red dwarfs, several ground based astrometry programs continued to do
the heavy lifting of measuring distances star-by-star (see, for
example the 38 references in \cite{Henry2018}, Table 4).  In one case,
a full-sky survey that provided thousands of parallaxes with a
particular focus on stars within 25 pc has been carried out by the
USNO \citep{Finch2016, Finch2018}.  Even so, most red dwarfs within 100
pc did not have distance measurements.

In 2013, the launch of the {\it Gaia} mission initiated a new era of
astrometry measurements for virtually all types of stars on the HRD
\citep{Gaia2016}.  In April 2018, {\it Gaia} Data Release 2 (DR2)
provided unprecedented sub-milliarcsecond parallax precision for over
a billion stars \citep{Gaia2018a} that can be utilized to study the
precise locations of individual stars on the HRD as well as
populations of stars.  The wealth of precise, all-sky data yields an
HRD that reveals never-before-seen features that were previously
impossible to identify.  For example, in \cite{Gaia-HRD} and
\cite{Kilic2018}, it was shown that (1) there are two main sequences
in the HRD for nearby stars caused by a metallicity difference of
about 1 dex, and (2) the distribution of nearby white dwarfs on the
HRD shows multiple sequences due to different atmosphere compositions
and mass distributions.

In this work, we present a new feature of the main sequence -- a gap
at mid-M dwarfs.

\section{The Gap}

\begin{figure}[ht]
\plotone{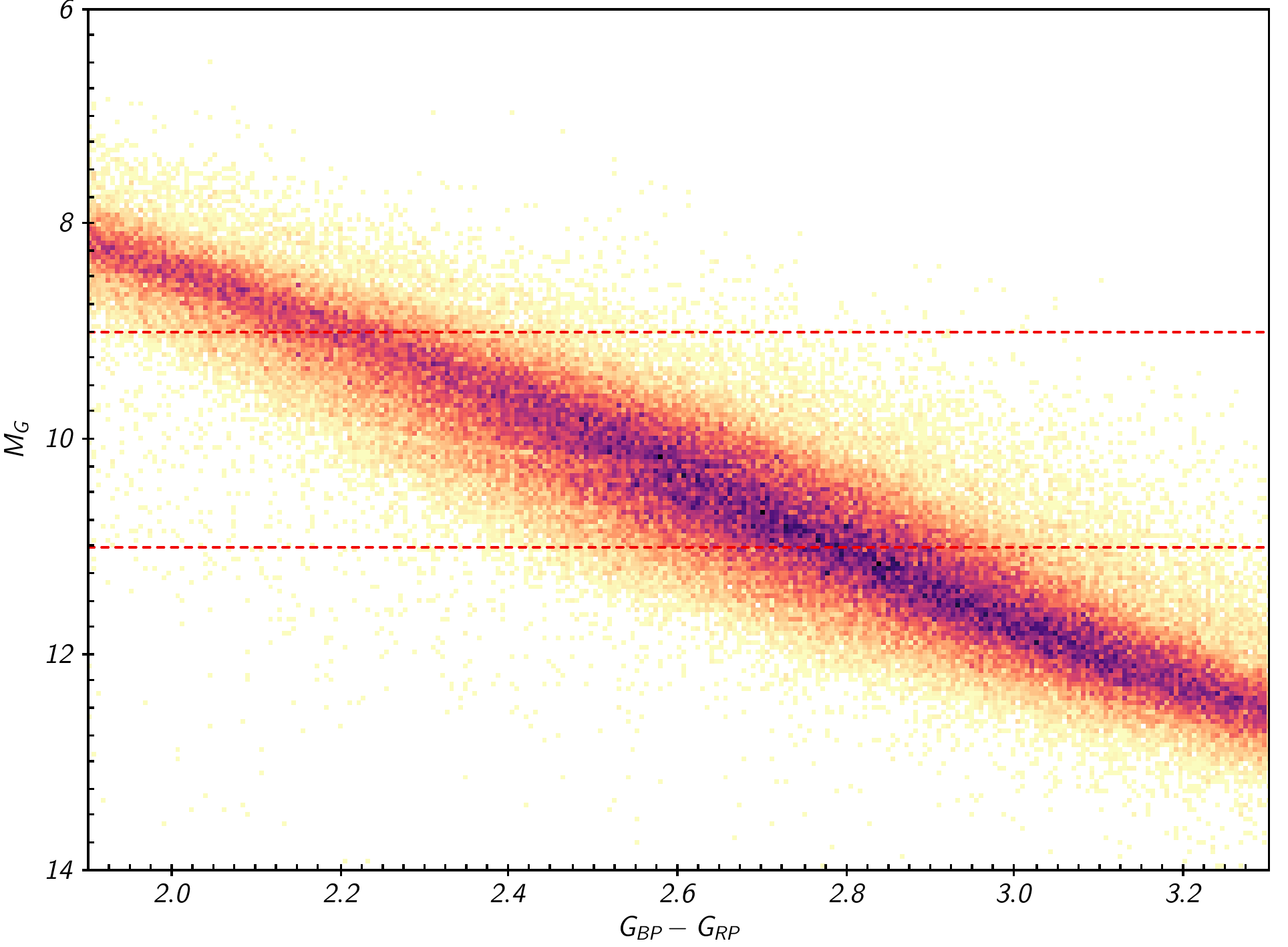}
\caption{A portion of the observational HRD for stars within 100 pc
  in the {\it Gaia} DR2 dataset, using $M_{G}$ and $G_{BP}-G_{RP}$.
  A thin, low density, gap is seen cutting through the main
  sequence. Two dashed lines ($M_{G}=$ 9 and 11) represent a region
  selected for further discussion, and plotted in Figure 2.}
\label{fig:HRD.zoom}
\end{figure}

A gap, shown in Figure~\ref{fig:HRD.zoom}, in the distribution of red
dwarfs is discovered using stars within 100 pc in DR2.  Data were
retrieved from the {\it Gaia} archive using an ADQL
script of~\\
\verb|SELECT * FROM gaiadr2.gaia_source WHERE parallax >=10|,
resulting in a sample of 700,055 extracted sources.  Data cleaning was
accomplished by using the three-step cuts discussed in
\cite{Lindegren2018}, primarily to remove low quality sources, many of
which fell between main sequence and white dwarfs on the HRD; a total
of 242,582 sources remained.  A portion of the HRD focused on M dwarfs
is shown in Figure~\ref{fig:HRD.zoom}, where a low density gap cuts
through the main sequence at $M_{G}\approx$10.

\begin{figure}[ht!]
\plottwo{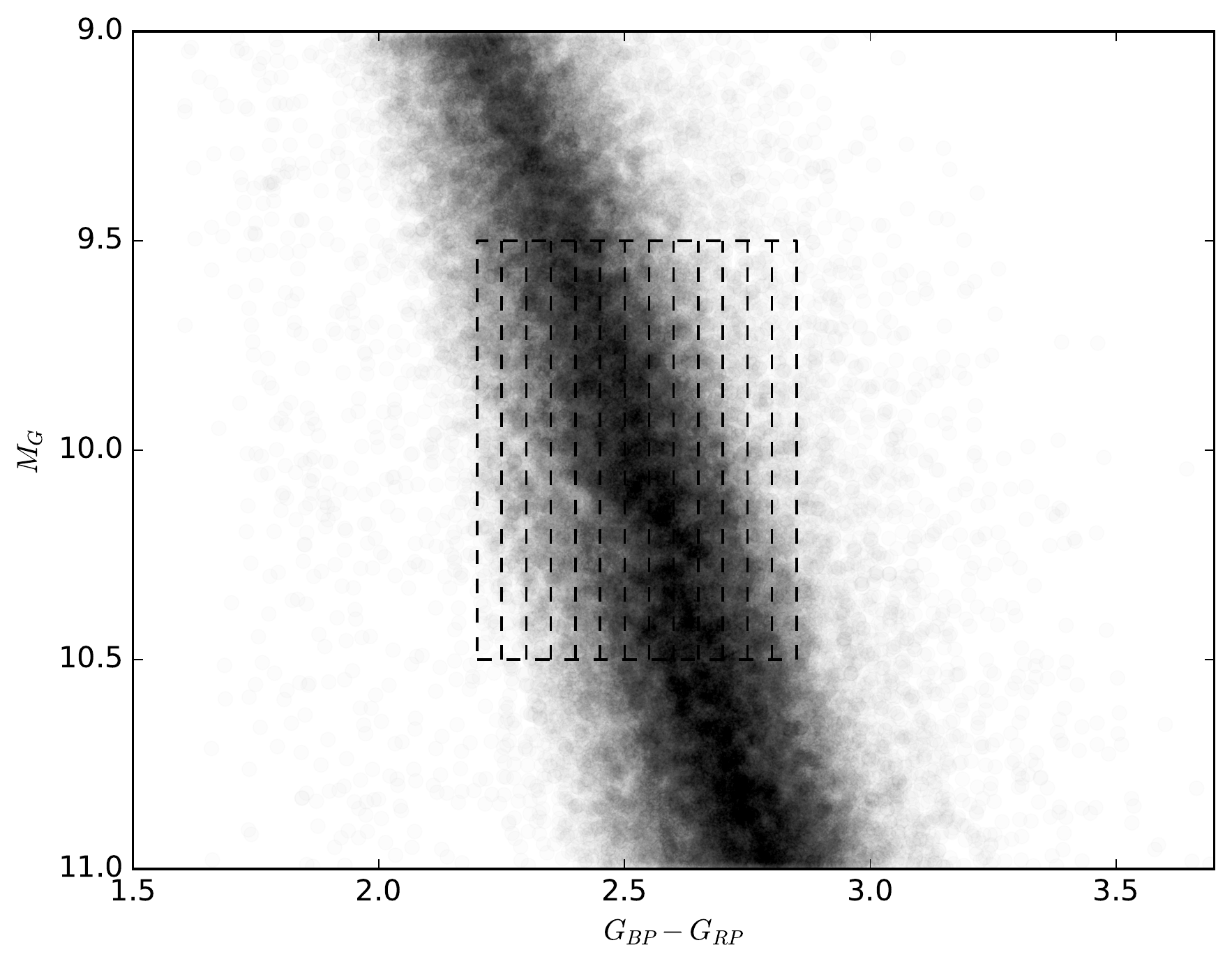}{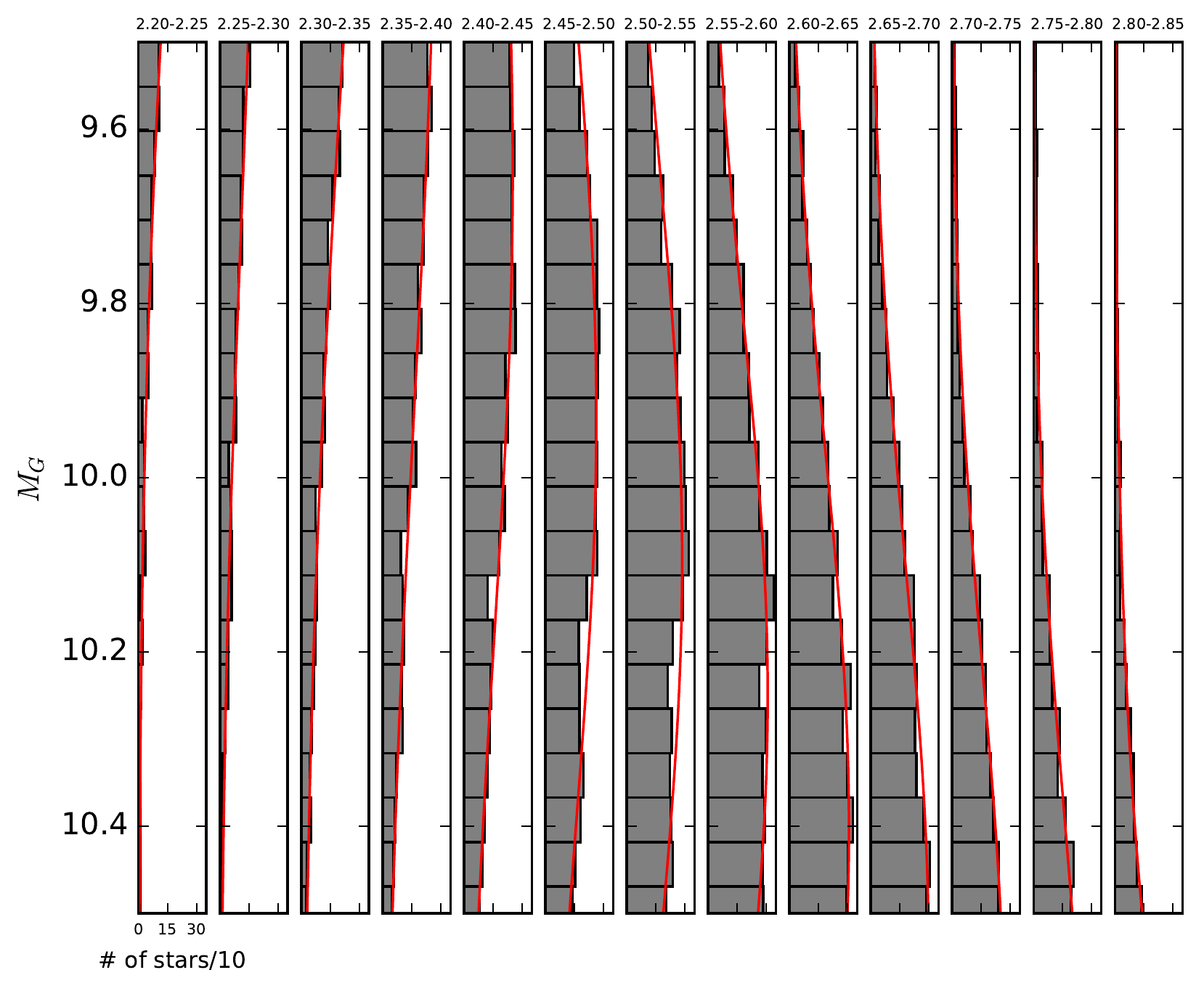}
\caption{(left) Sliced strips of the main sequence near the gap are
  shown, with vertical cuts in $G_{BP}-G_{RP}$ color.  (right)
  Distributions of the number of stars in each $G_{BP}-G_{RP}$ strip
  are shown, with Gaussian fits outlined in red.  Each vertical
  strip is 0.05 mag wide in color, and the histogram bin sizes are
  0.051 mag in $M_G$ to optimize the gap's effect.}
\label{fig:slice}
\end{figure}

To further highlight and quantify the stellar population near this
gap, a zoomed region between the two dashed lines in
Figure~\ref{fig:HRD.zoom} has been sliced into several strips as shown
in the left panel of Figure~\ref{fig:slice}.  Histograms of star
counts in each strip are illustrated in the right panel and show that
the gap progresses from $M_{G}\sim$10.09 in the $G_{BP}-G_{RP}$=
2.35--2.40 bin to $\sim$10.24 in the 2.55--2.60 bin.  Comparison to
Gaussian models of star distributions along the $M_{G}$ axis (shown in
red in the right panel of Figure~\ref{fig:slice}) indicates that the
decrements of stars in the gap regions are $\sim$24\% and 14\% less
than the Gaussian curves, respectively, in these two bins.  This also
shows the gap is very narrow, with a width of only about 0.05
mag. Furthermore, it is more pronounced at blue colors than at
red. The largest decrement is 28\% in the 2.40--2.45 bin. The details
of these decrements in each bin are summarized in
Table~\ref{tbl:percentage}.

\begin{deluxetable}{ccccc}[h]
\tablecaption{Percentage decrement along the gap for different samples\label{tbl:percentage}}
\tablehead{
\colhead{} &
\colhead{} &
\colhead{0-100pc} &
\colhead{100-130pc} &
\colhead{120-130pc} \\
\colhead{}  &
\colhead{}  &
\colhead{70,700} &
\colhead{81,600} &
\colhead{31,610}\\
\colhead{$G_{BP}-G_{RP}$ bin}                       &
\colhead{$M_{G}$ with largest decrement}    &
\colhead{percentage decrement} &
\colhead{percentage decrement} &
\colhead{percentage decrement}
     }
\colnumbers
\startdata
2.30-2.35    &   10.04   &           19(73/90.2)     &  11(90/100.6)   & 25(29/38.4)\\  
2.35-2.40    &   10.09   &           24(95/125.3)   &  12(125/142.2) & 21(42/52.9)\\   
2.40-2.45    &   10.14   &           28(122/168.8) &  19(152/188.6) & 24(53/69.6)\\   
2.45-2.50    &   10.19   &           23(173/225.5) &  30(178/254.5) & 28(70/97.4)\\   
2.50-2.60    &   10.24   &           22(211/271.4) &  16(256/302.9) & 10(107/118.9)\\   
2.55-2.60    &   10.24   &           14(264/307.7) &  16(290/344.2) & 18(108/131.5)\\   
2.60-2.65    &   10.29   &             7(275/297.0) &  11(303/340.1) & 12(121/137.9)\\  
2.65-2.70    &   10.34   &           12(236/296.3) &  12(271/307.4) & 11(107/120.5)\\  
\enddata
\tablecomments{The total numbers of stars in each sample are given in
  the second row, and these stars have $M_{G}$ between 9.0 and
  11.0. Two numbers in the parentheses indicate the actual start
  counts (integer) and fitted Gaussian counts (real number) in each
  bin.}
\end{deluxetable}

\section{The Gap in Different Colors and More Distant Samples}

\begin{figure}[ht!]
\plotone{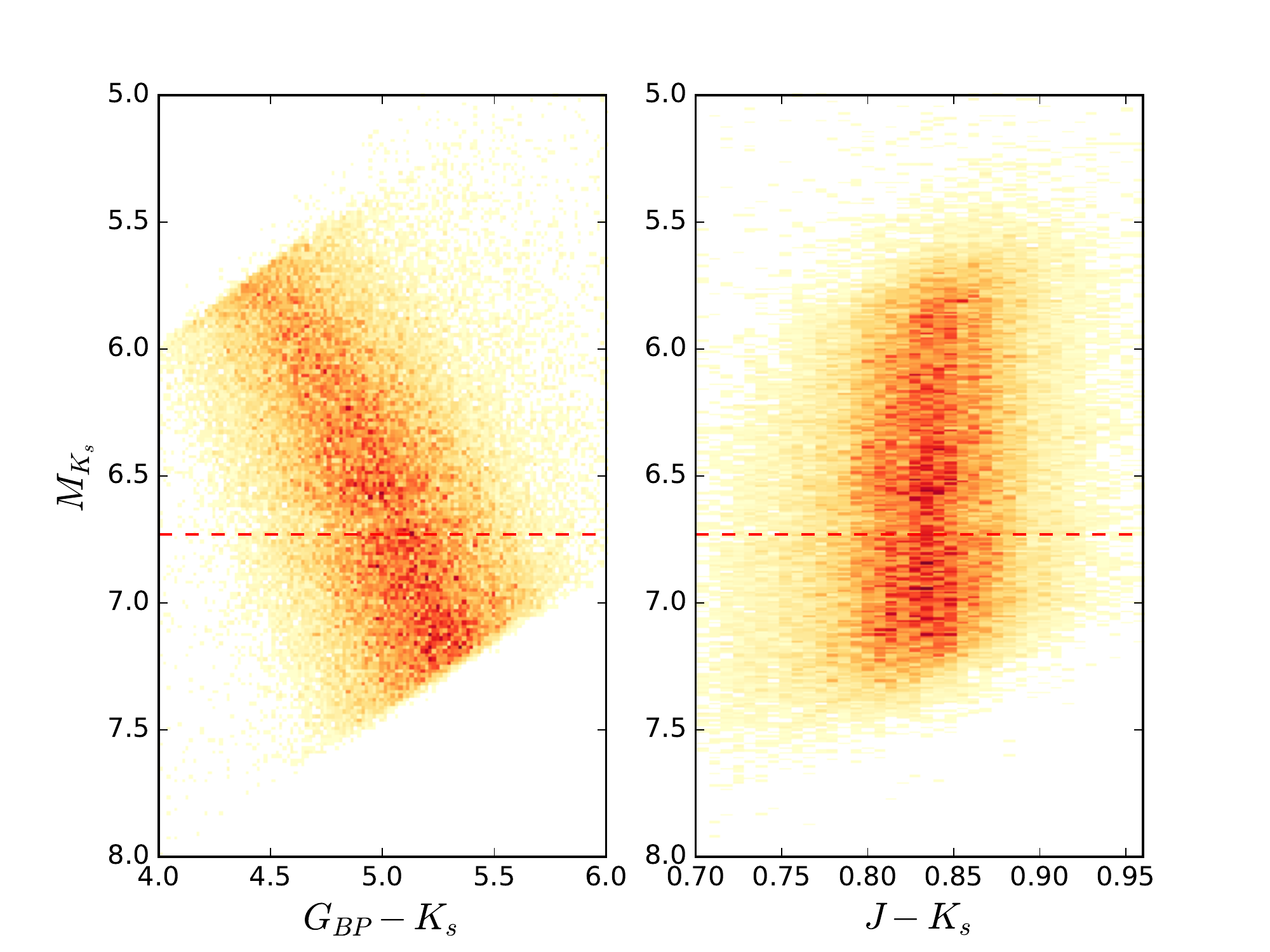}
\caption{Two observational HRDs are shown using $M_{K_{s}}$ for
  luminosity and $G_{BP}-K_{s}$ and $J-K_{s}$ for temperature.  The
  gap is seen in both plots, indicating that the gap persists at both
  optical and infrared wavelengths.  Red dashed lines mark the
  $M_{K_{s}}=$6.73 value that corresponds to stars with mass
  0.35$M_{\odot}$ based on the mass-luminosity relation of
  \cite{Benedict2016}. The gap is seen in both panels
    slightly above the dashed lines.}
\label{fig:twoHRD}
\end{figure}

An important consideration is whether or not the gap persists in
colors other than $G_{BP}-G_{RP}$.  To investigate, 70,700 stars
between the two dashed lines in Figure~\ref{fig:HRD.zoom} were
cross-matched against the 2MASS catalog.  Because many of these nearby
red dwarfs have high proper motions, the J2015.5 coordinates in DR2
were adjusted to J2000.0 using the DR2 proper motions so that their
coordinates would be close to their positions in 2MASS images, which
were taken from 1998--2001.  A 5\arcsec~search radius was then used to
find matches of DR2 sources to 2MASS sources.  Figure~\ref{fig:twoHRD}
shows two different HRDs, $G_{BP}-K_{s}$ vs $M_{K_s}$ and $J-K_{s}$
vs.~$M_{K_s}$, that illustrate the results.  Somewhat surprisingly,
the gap is evident in these two observational HRDs as well.  {\it
  Thus, the gap is not unique to the {\it Gaia} photometry, and is not
  caused by specific stellar spectroscopic features in the optical or
  near-infrared bands alone.}

\begin{figure}[ht!]
\plotone{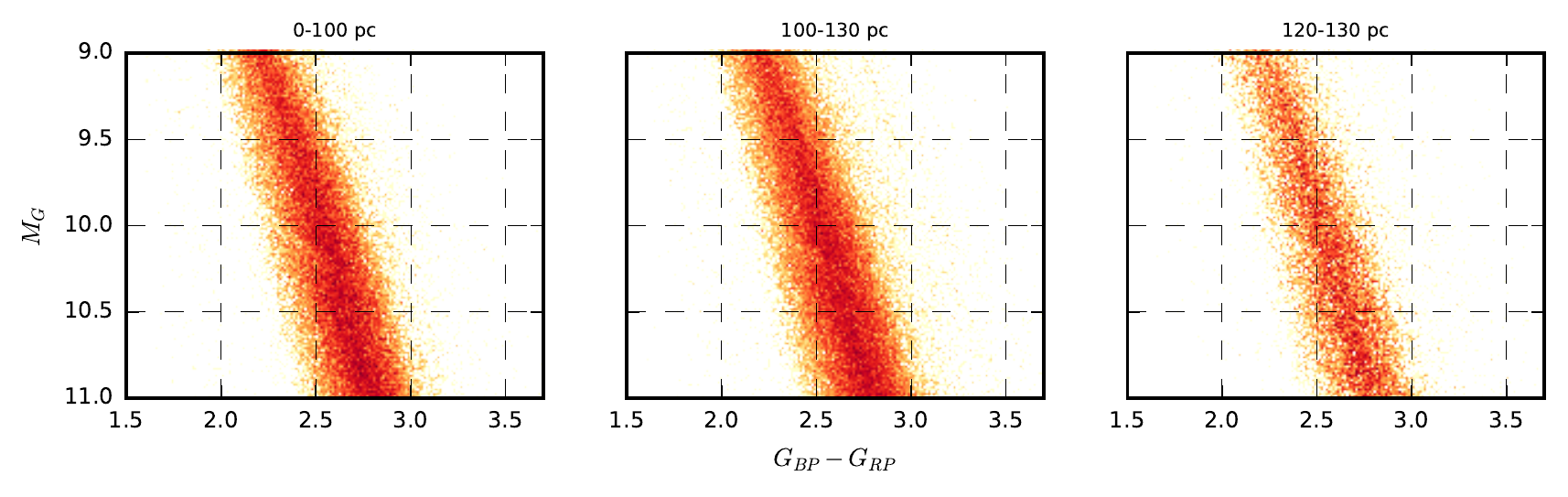}
\caption{Three HRDs of different samples, 0--100, 100--130 and
  120--130 pc from {\it Gaia} DR2 are shown from left to right. The
  total numbers of stars used in each sample are given in
  Table~\ref{tbl:percentage}. }
\label{fig:3panel}
\end{figure}

In order to test the persistence of the gap, we extracted two other
samples from DR2 using stars in the 100--130 pc and 120--130 pc
shells. After applying the three-step cuts in \cite{Lindegren2018},
81,600 and 31,610 stars are left, respectively. The 100--130 pc shell
is selected to have a similar number of stars and volume as stars in
0--100 pc.  The other 120--130 pc shell contains about half the number
of stars in 0--100pc and a totally different set of stars. HRDs for
the three samples are presented side-by-side in
Figure~\ref{fig:3panel}.

We see the same gap at the exact same location in all three HRD.  The
120--130 pc sample size is only $\sim$45\% the size of the sample in
the 0--100 pc range, so the gap is not as obvious. Although the
100--130 pc sample size is the largest, its gap is not the sharpest
because stars in 100--130 pc range have relatively smaller parallax to
parallax error ratios. Hence, parallax errors become more significant
at larger distances, thereby shifting stars slightly on the HRD, with
a consequent blurring of the distribution near and in the gap.  We
note that samples of the nearest stars, such as those within 25 pc in
DR2, do not show the gap because there are only $\sim$4000 points
spanning the entire main sequence.

\section{Discussion}

The detection of a new feature in the HRD implies a previously unknown
characteristic of stars that may be linked to fundamental
astrophysics.  This leads to a number of questions that need to be
considered.

{\bf Is the gap due to a bias in {\it Gaia} data?} According to
\cite{Gaia2018a}, (1) ``a fraction of the bright stars at $G<$7 is
still missing with no stars brighter than $G=$1.7 mag appearing in
DR2'' and (2) 17\% of stars with proper motions greater than 0\farcs6
yr$^{-1}$ could still be missing.  Thus, there appear to be no biases
that would omit stars near the gap in the DR2 dataset.

{\bf Is this gap caused by the presence of multiple main sequences?}
The double main sequence bands shown in Figure 21 of \cite{Gaia-HRD}
are almost parallel and almost evenly spread apart.  Globular clusters
like NGC6397 \citep{Milone2012} and $\omega$ Centauri
\citep{Bedin2004} also have double main sequences and their dividing
lines are similar to the one seen in \cite{Gaia-HRD}.  The gap we
outline here for nearby stars has a very different slope, so is not
consistent with gaps resulting from multiple main sequence
populations.

{\bf Can the gap be reproduced for stars in other parts of the
  Galaxy?}  The ideal laboratories in which to search for the gap are
open or globular clusters, because such stellar groups comprise
relatively clean samples given the similar ages and metallicities of
members.  Unfortunately, no open clusters are known to have the tens
of thousands (at least) of stars needed to define the gap.  The
nearest globular cluster, NGC 6397, has been observed by {\it Gaia},
but cannot be used to confirm the gap because the faintest stars
observed are $\sim$1 magnitude bluer in $G_{BP}-G_{RP}$ color than the
gap shown in Figure~\ref{fig:HRD.zoom}.  \cite{Richer2008} presented a
color-magnitude diagram for NGC 6397 from a deep survey using {\it
  HST/ACS} that reached down the main sequence to late-type M dwarfs
(see their Figure 6).  Unfortunately, because of the low number of
stars around the gap region, no reliable test of the gap can be done.

{\bf Why has this gap not been seen before?} Before the DR2 dataset,
the available number of parallaxes for M dwarfs were both (1) limited
in number, and (2) had larger errors.  Because the gap is narrow and
subtle, these two attributes of the available parallax sets precluded
identifying the gap --- a much larger set of high-precision parallaxes
is needed.

\begin{figure}[ht!]
\plotone{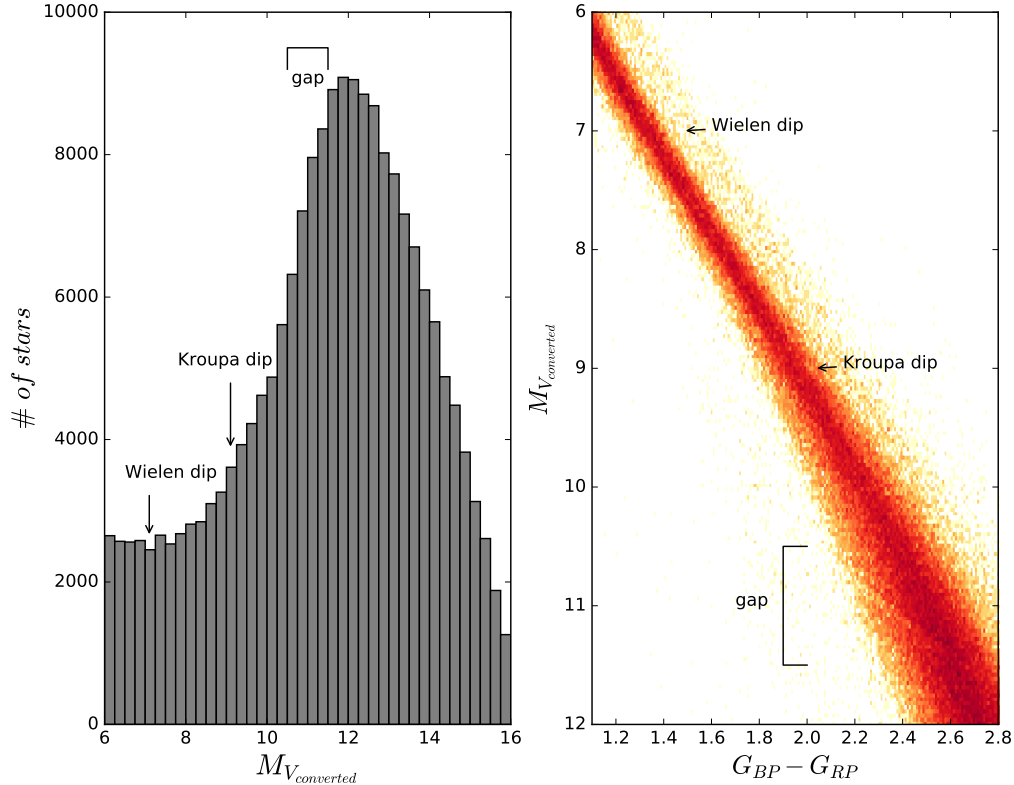}
\caption{(Left) The stellar luminosity function based on {\it Gaia}
  DR2 data is shown.  The Johnson
  $V$ magnitudes have been converted using the transformation
  discussed in Appendix~\ref{sec:app1}.  The ``Wielen dip'' and
  ``Kroupa dip'' are roughly marked at $M_{V}\approx7$ and
  $9$, respectively.  The gap discussed here is outlined with a
  bracket shown at
  $M_{V}$=10.5--11.5. (Right) The different dips and the DR2 gap are
  marked on the HRD, and this DR2 gap is very obvious.}
\label{fig:LF}
\end{figure}

\begin{figure}[ht!]
\plotone{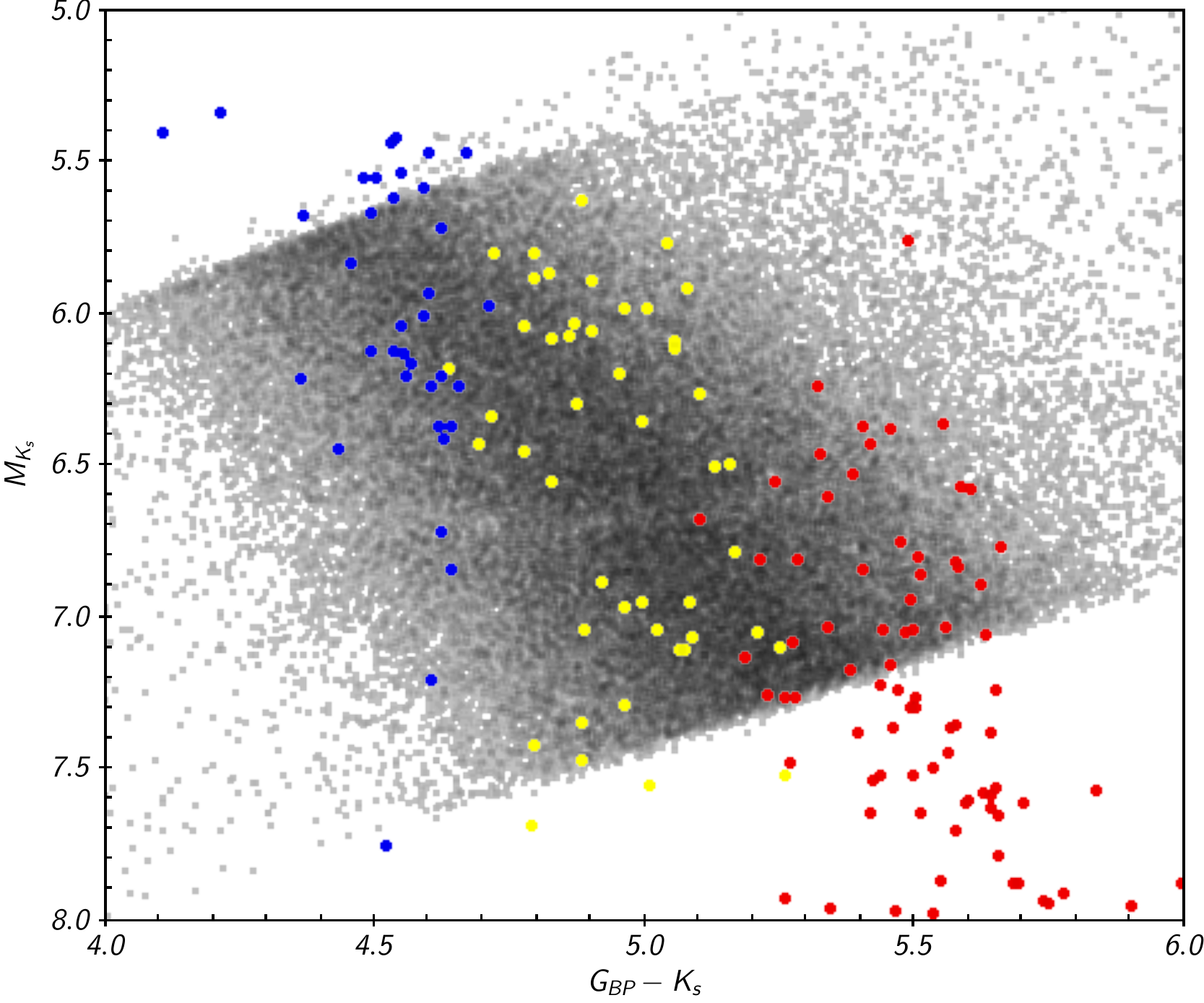}
\caption{An observational HRD using $G_{BP}-K_{s}$ and $M_{K_{s}}$
  to separate different spectral types of stars.  Blue, yellow and
  red circles represent known single stars within 25 pc with
  spectral types of M2.0V, M3.0V, and M4.0V, respectively. }
\label{fig:M234}
\end{figure}

{\bf Is the gap related to the ``Wielen dip'' or ``Kroupa dip''?}  The
Wielen dip was first reported in \cite{Wielen1983} while studying the
luminosity function of nearby stars using the 1969 version of the
Catalogue of Nearby Stars \citep{Gliese1969}.  The dip occurs at
$M_{V}\approx$7, corresponding to mid-K type dwarfs, i.e., at much
earlier types than the gap described here near M3.0V type
(Figure~\ref{fig:M234}).  \cite{Kroupa1990} and \cite{Kroupa2002}
proposed that this depression/plateau in the luminosity function is
due to H$^{-}$ opacity becoming increasingly important in the
atmospheres of K dwarfs, reducing their luminosities to create the
(slight) dip marked in Figure~\ref{fig:LF}\footnote{This luminosity
  function has been generated using DR2 data and is only used as a
  demonstration.  Generating the most accurate luminosity function
  based on DR2 is beyond the scope of this work and would, in
  particular, require the careful deconvolution of multiple systems
  resolved by {\it Gaia} or not.  Nonetheless, the overall shape of
  the luminosity function is still correct.}.

The other so-called ``Kroupa dip'' at $M_{V}\approx$9 can be seen in
the stellar luminosity function of field stars in \cite{Kroupa1990}
and its location is marked in
Figure~\ref{fig:LF}. \cite{Elsanhoury2011} studied both dips using
several open clusters, and showed ``both dips may not be strongly
visible'' in the luminosity function of clusters. Even though both
features are called ``dips'', they are almost invisible in the HRD
plot of Figure~\ref{fig:LF} compared to this DR2 gap.

The gap discussed here spreads from $M_{Vconverted}$\footnote{See
  Appendix~\ref{sec:app1} for derivation methodology for
  $M_{Vconverted}$.}$\sim$10.5--11.5 and is dependent on color, so is
not clearly represented in the histogram of Figure~\ref{fig:LF}.
Regardless, it is clear that both Wielen and Kroupa dips are much
bluer than the gap we see in DR2 data.

{\bf Is the gap related to the onset of full convection in $\sim$M3.0V
  stars?} As shown in Figure~\ref{fig:M234}, the gap maps onto
spectral type $\sim$M3.0V, where blue, yellow, and red points
represent single stars within 25 pc with spectral types of M2.0V,
M3.0V, and M4.0V, respectively.  This is the location on the main
sequence where interior structure models show a transition from
partially to fully convective stars \citep{Limber1958, Hayashi1963,
  Dorman1989, Chabrier1997} at $\sim$0.35$M_{\odot}$.  Recent
theoretical stellar models \citep{Marigo2017, Baraffe2015, Dotter2008}
are generally smooth from high to low masses, including through the M
dwarfs, with no abrupt transitions.  

Model isochrones can help us estimate the mass and radii associated
with the gap, although we caution that stellar models still have
difficulties in matching the color-magnitude diagrams of low mass
stars in clusters \citep{Chen2014}. Two selected
  models to be compared to the {\it Gaia} data are PARSEC
  \citep{Marigo2017} and YaPSI \citep{Spada2017}.

\begin{figure}[ht!]
\plotone{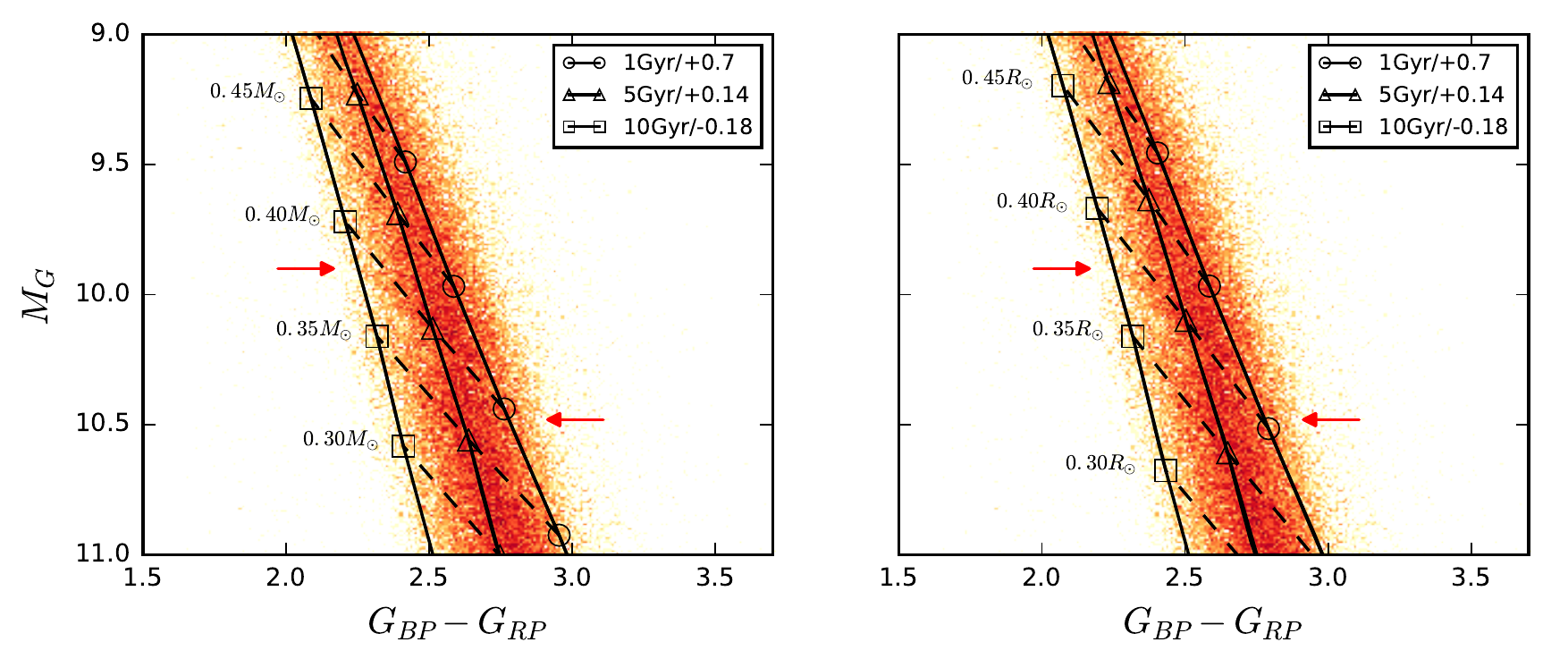}
\caption{Three different isochrones (solid lines) from PARSEC are
  shown on the HRD. Circles, triangles and square boxes represent
  three different isochrones given in the legends. Dashed lines
  represent equal masses (left figure) and radii (right figure) on all
  three isochrones. Because these two figures are crowded, identifying
  this gap becomes difficult. Two arrows mark the gap location and the
  gap feature appears to have a shallower slope than either the
  equal-mass line at 0.4$M_{\odot}$ or equal-radius line at
  0.4$R_{\odot}$. }
\label{fig:iso}
\end{figure}

We plot three different isochrones from PARSEC\footnote{Isochrones are
  retrieved from \url{http://stev.oapd.inaf.it/cgi-bin/cmd}.}  in
Figure~\ref{fig:iso}. Most isochrone parameters are default values at
this website, other than ages and metallicities. Three isochrones
shown are 1 Gyr/[M/H]=+0.7, 5 Gyrs/[M/H]=+0.14 and 10
Gyrs/[M/H]=$-$0.18. We can see that the gap intercepts the
0.4$M_{\odot}$ equal-mass and 0.4$R_{\odot}$ equal-radius
lines. Although the gap's slope is somewhat different than these two
lines, we expect the slope of the gap is caused by varying
metallicities and ages of these nearby stars at a given mass or
radius. As Figure~\ref{fig:HRD.zoom} shows, this gap is more prominent
in the blue, implying that low metallicity stars tend to deepen this
gap.

\begin{figure}[ht!]
\plotone{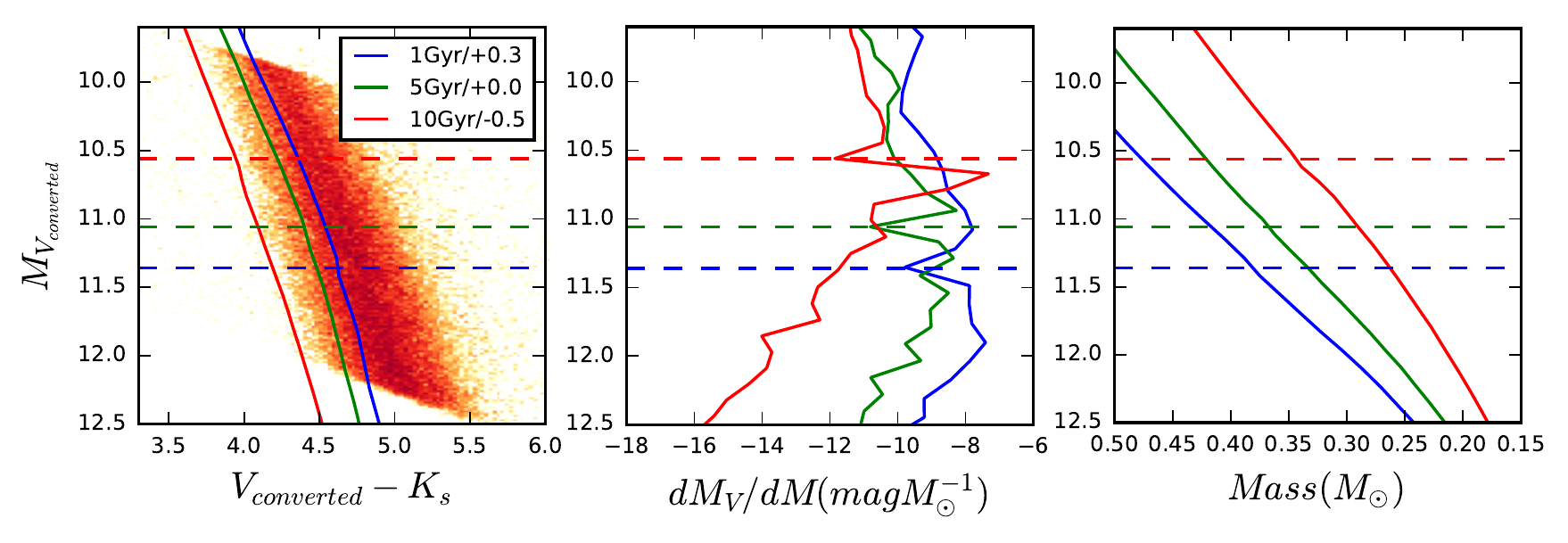}
\caption{Three different isochrones (solid lines) from
    YaPSI are plotted against the {\it Gaia} data on the HRD.  Blue,
    green and red lines represent three different isochrones,
    1Gyr/[Fe/H]=+0.3/X=0.717/Z=0.03, 5Gyr/[Fe/H]=+0.0/X=0.733/Z=0.016
    and 10Gyr/[Fe/H]=-0.5/X=0.744/Z=0.005, respectively, and all
    isochrones have Y=0.25. The {\it Gaia} BP magnitudes are converted
    to $V$ using the relation discussed in Appendix~\ref{sec:app1}. We
    note that these isochrones do not contain 2MASS $K_s$ magnitudes,
    but we believe the difference between $K$ and $K_s$ is small. The
    middle plot is the slope of the mass-luminosity relations of these
    three metallicities shown on the right. Three colored dashed lines
    mark where the MLR slopes have local minima at $M_{V}=$11.36,
    11.06 and 10.56.}
\label{fig:yapsi}
\end{figure}

We also retrieved three different YaPSI isochrones
  with $VRIJHK$ colors based on \cite{Worthey2011} from their website
  \footnote{\url{http://www.astro.yale.edu/yapsi/download_grids.html}}. The
  three different isochrones are shown in the left plot of
  Figure~\ref{fig:yapsi} along with the {\it Gaia} data. Because YaPSI
  models have a much finer grid step size (0.01M$_\odot$) than those
  in PARSEC (0.05$M_\odot$), a discontinuity in the slope of the
  isochrone is revealed near the location of the gap. As
  \cite{Spada2017} pointed out, the kink close to 0.35M$_\odot$ in the
  mass-luminosity relation occurs because of the transition from fully
  convective to solar-like interior structure.  These kinks are where
  the slopes of mass-luminosity relations change and are marked as
  three dashed lines in Figure~\ref{fig:yapsi}. The transitions are
  approximately at 0.38, 0.37, and 0.33M$_\odot$, respectively for
  [Fe/H]=+0.3, 0.0 and $-$0.5. Discontinuities in the slopes of the
  isochrones are associated with a drop in the number densities in the
  gap region.  We expect that the increment in stellar number $N$
  varies as
  $${{dN}\over{dM_V}} = {{dN}\over{dM}} / {{dM_V}\over{dM}}$$ where
  ${{dN}\over{dM}}$ is the continuous stellar mass function and
  ${{dM_V}\over{dM}}$ is the predicted change in absolute magnitude
  with stellar mass.  The right panel of Figure 8 shows the
  discontinuities in the $(M, M_V)$ plane, while the central panel
  shows the absolute magnitude as a function of the local numerical
  derivative ${{dM_V}\over{dM}}$.  We see that the predicted
  magnitudes where the model ${{dM_V}\over{dM}}$ attains a local
  minimum correspond to those magnitudes with fewer stars, i.e., where
  the gaps are located on the HRD.  We note that the three YaPSI
  isochrones are significantly shifted to the blue relative to the
  {\it Gaia} data, unlike the PARSEC models shown in
  Figure~\ref{fig:iso}.

Observationally, the empirical mass-luminosity relation
\citep{Benedict2016}, temperature-radius relation \citep{Ribas2006,
  Boyajian2012}, and large scale magnetic topology studies of M dwarfs
\citep{Morin2010} also show relatively smooth transitions based on
small samples.  The narrow gap implies a ``sudden'' transition from
one state to another, which happens to fall near the mass where stars
are modelled to become fully convective.  This presents a conundrum:
we typically refer to stars beginning to be fully convective at a
given temperature corresponding to spectral type $\sim$M3.0V, implying
a vertical gap in the HRD, but we observe that the gap is roughly
horizontal in $M_G$, $M_V$ and $M_{Ks}$.

\begin{figure}[ht!]
\plotone{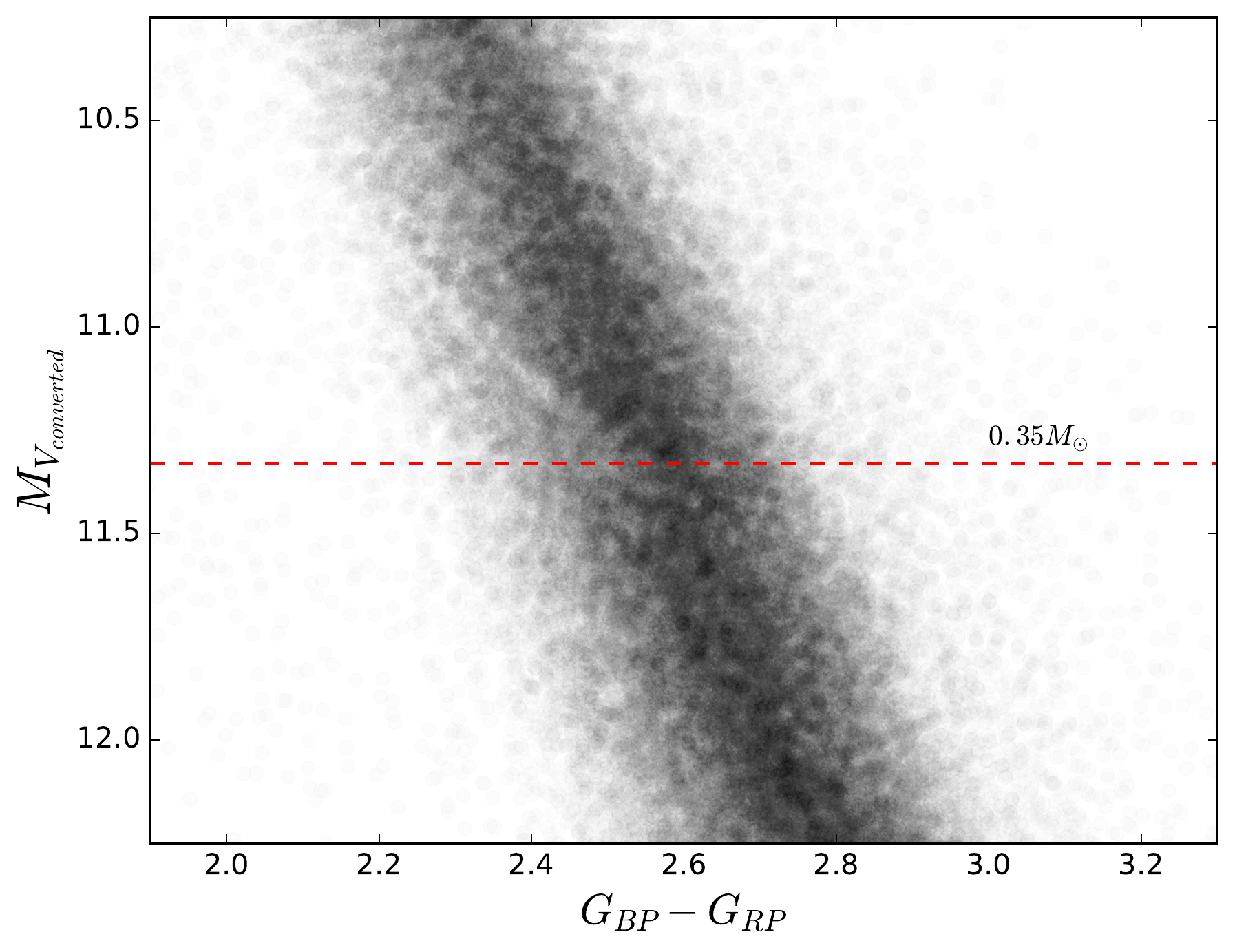}
\caption{An observational HRD using $G_{BP}-G_{RP}$ and
  $M_{V_{converted}}$ is shown.  The red dashed line that intersects
  the gap marks the $M_{V_{s}}=$11.33 value that corresponds to stars
  with mass 0.35$M_{\odot}$ based on the mass-luminosity relation of
  \cite{Benedict2016} from nearby stars with various metallicities.}
\label{fig:M35}
\end{figure}

Using the mass-luminosity relation of \cite{Benedict2016},
0.35$M_{\odot}$ corresponds to $M_{K_{s}}$ = 6.73 and $M_{V}$ = 11.33.
These values are marked with dashed lines in Figure~\ref{fig:twoHRD}
for $M_{K_{s}}$ and Figure~\ref{fig:M35} for $M_{V}$, and in both
cases fall near the gaps.  It is important to point out that the
0.35$M_{\odot}$ value for the transition to fully convective stars is
approximate, and the \cite{Benedict2016} sample used to estimate the
absolute magnitudes for 0.35$M_{\odot}$ stars is a heterogeneous mix
of stars in the solar neighborhood with various metallicities and
magnetic properties.

All the evidence shown above is quite compelling --- it is probable
that gap and the transition to fully convective stars are linked. More
importantly, no other region of the main-sequence has a similar gap
like this. The result implies that stars above the gap may prove to be
mostly convective with a thin radiative layer and more massive, while
those below are fully convective less massive and smaller.

\section{Future Study of the Gap}

This new feature on the HRD shows the power of high precision
astrometry and challenges how we understand all regions in the
fundamental map of stellar luminosities and temperatures.  It appears
that there is a small slice of the HRD that is less populated, at
least relative to surrounding regions in luminosity and temperature.
Based on the evidence presented here, the observed drop in stellar
numbers in the gaps is probably related to a subtle change in
structure (decrease in radius) that occurs at the transition between
partially and fully convective stars.

We suggest the following studies be undertaken to further understand
this new feature on the HRD.  (1) Examine the stars currently falling
in the gap to evaluate contamination from unresolved multiples,
thereby providing a better assessment of the true depth of the
decrement in population.  (2) Measure accurate dynamical masses, radii
and metallicities for stars on both sides of the gap, as well as those
remaining in the gap, to understand the characteristics across this
region of the HRD.  (3) Determine the rotational periods for stars in
and around the gap to see if there are any trends.  (4) Evaluate the
photometric variability as a proxy for investigating the effects of
magnetism, or measure magnetic topology of stars in this region of the
HRD to determine if there are any variations that provide clues as to
the cause of the gap.

\acknowledgments

We would like to thank Frederic Arenou for helpful discussions about
DR2 data and the anonymous referee for very useful comments. This work
is supported by the National Science Foundation under grant
AST-1715551. This work has made use of data from the European Space
Agency (ESA) mission {\it Gaia}
(\url{https://www.cosmos.esa.int/gaia}), processed by the {\it Gaia}
Data Processing and Analysis Consortium (DPAC,
\url{https://www.cosmos.esa.int/web/gaia/dpac/consortium}).  Funding
for the DPAC has been provided by national institutions, in particular
the institutions participating in the {\it Gaia} Multilateral
Agreement.  This research has made use of the SIMBAD database,
operated at CDS, Strasbourg, France. This research has made use of
NASA's Astrophysics Data System. This publication makes use of data
products from the Two Micron All Sky Survey, which is a joint project
of the University of Massachusetts and the Infrared Processing and
Analysis Center/California Institute of Technology, funded by the
National Aeronautics and Space Administration and the National Science
Foundation.

%



\software{TOPCAT \citep{TOPCAT}}

\appendix

\section{Converting {\it Gaia} ($G_{BP}$ and $G_{RP}$) to
  Johnson--Kron--Cousin ($V$ and $I$)}
\label{sec:app1}

\begin{figure}[ht!]
\plotone{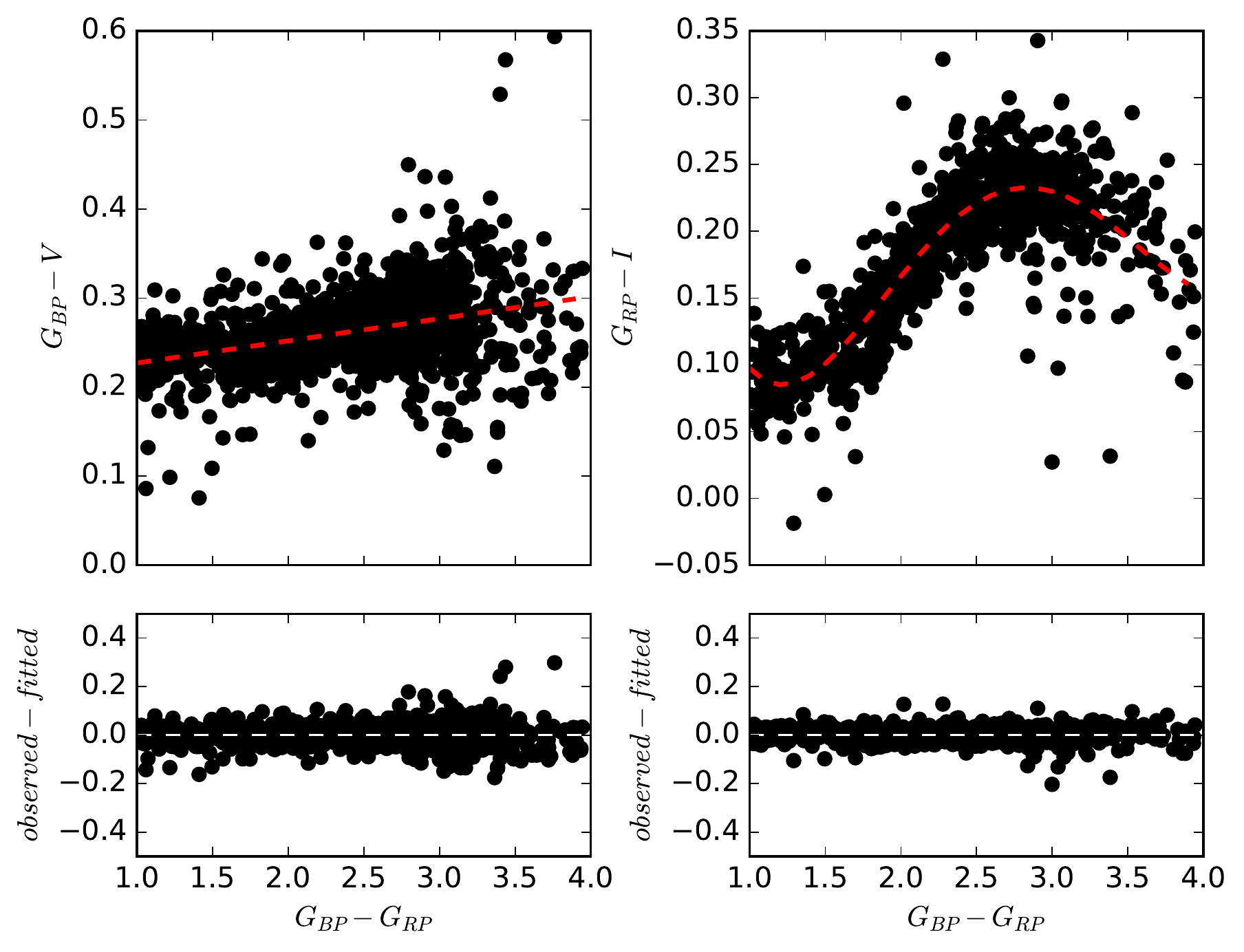}
\caption{The top left panel represents a relation between
  $G_{BP}-G_{RP}$ and $G_{BP}-V$, where the lower left panel shows
  the differences between the observed and fitted values.  The red
  dashed line in the top figure represents the best fit line, and a
  white dashed line in the bottom figure represents zero point.  The
  right two figures are similar, but for the $G_{BP}-G_{RP}$ and
  $G_{RP}-I$ relation.}
\label{fig:convert}
\end{figure}

\cite{Evans2018} provides a transformation equation to convert {\it
  Gaia} $G$ magnitudes to Johnson $V$.  Because the $G$ bandpass is
much wider than $V$ and includes significantly more red flux, we
provide transformations from $G_{BP}$ to Johnson $V$ and $G_{RP}$ to
Kron-Cousin $I$.  We use ground-based photometry measurements for
1,223 single stars with $V$ for the $G_{BP}$--$V$ relation and 984
single stars with $I$ for the $G_{RP}$--I relation.  The stars have
spectral types K0V to M6V and are within 25 pc.  The total numbers of
stars used for these two relations are different because some stars do
not have both $V$ and $I$ magnitudes.  The two relations are shown in
Figure~\ref{fig:convert}, and transformation coefficients are given in
Table~\ref{tbl:coeff}.

\begin{deluxetable}{ccccccc}
\tablecaption{Coefficients of the polynomials used to convert photometric colors \label{tbl:coeff}}
\tablehead{
\colhead{conv} & 
\colhead{$c_{0}$} & 
\colhead{$c_{1}$} &
\colhead{$c_{2}$} & 
\colhead{$c_{3}$} & 
\colhead{$c_{4}$} &
\colhead{$\sigma$}
}
\colnumbers
\startdata
$G_{BP}-V$   &  0.20220  &  0.02489   &   \nodata   &  \nodata    &
\nodata  & 0.04035 \\
$G_{RP}-I$    &  0.75319  &  -1.41105  &  1.00136   &  -0.27106  &
0.02489 & 0.02770
\enddata
\tablecomments{These relations are valid for
  1.0$\leq$$G_{BP}-G_{RP}$$\leq$4.0.  The transformations have the
  following format where color=$G_{BP}-G_{RP}$:
  $conv=c_{0}+c_{1} \times color + c_{2} \times color^{2}+ c_{3}
  \times color^{3} +c_{4} \times color^{4}$. The seventh column,
  $\sigma$, represents standard deviations of differences between
  observed and fitted values shown in the bottom two panels of
  Figure~\ref{fig:convert}.}
\end{deluxetable}

\section{Plotting programs used to show this gap}
Because this gap is not obvious on the HRD, we use two different plotting
programs and settings to show this gap and they are summarized below.

\noindent Fig 1: TOPCAT/density plot\\
Fig 2 and 9: Python Matplotlib/scatter plot with alpha=0.01\\
Fig 3: Python Matplotlib/2D histogram with a regular color scale \\
Fig 4, 5, 7 and 8: Python Matplotlib/2D histogram with a logarithmic color scale\\
Fig 6: TOPCAT/scatter plot



\end{document}